# Type-I and type-II Dirac fermions in graphene with nitrogen line defects

Honghong Zhang, Yuee Xie, Chengyong Zhong, Zhongwei Zhang and Yuanping Chen[*]

*School of Physics and Optoelectronics, Xiangtan University, Xiangtan, Hunan, 411105, China.*

**Abstract:** Recently, type-II Dirac fermions characterized by strongly tilted Dirac cones have been proposed. The new fermions exhibit unique physical properties different from the type-I Dirac fermions in graphene, and thus attract tremendous attentions. Up to date, all type-II fermions are only found in the heavy compounds with strong spin obit coupling. Here, we propose that both type-I and type-II Dirac fermions can exist in the graphene embedding nitrogen line defects. While the types of Dirac fermions are determined by the size *W* of graphene nanoribbons between the line defects. By comparing the two types of Dirac fermions, their different physical properties and originations are revealed directly. Remarkably, the type-I Dirac points induce one Fermi arc corresponding to edge states along the armchair direction, while the type-II Dirac points induce two Fermi arcs corresponding to two sets of edge states along the zigzag direction. These results not only expand our views on the Dirac fermions in two-dimensional structures, but also extend their applications in electronics.

---

[*]Corresponding author. E-mail address: chenyp@xtu.edu.cn (Y. Chen).

# 1. Introduction

Since 2004, graphene has been extensive studied because of its excellent electronic properties and extensive applications [1-3]. Most unique electronic properties of graphene are originated from its special band structure with Dirac cones [4-10]. Therefore, exploring physical properties of Dirac fermions and finding related materials have been an advanced field in condensed matter physics. Recently, a new type of Dirac fermions has been proposed and attracted tremendous attentions. It is characterized by strongly titled Dirac cones, and thus emerge at the boundary between electron and hole pockets. The new fermions are named as type-II Dirac fermions, in contrast to the type-I Dirac fermions in graphene [11, 12]. The two types of Dirac fermions exhibit very different physical properties. For example, the materials possessing type-I fermions are semimetals because the Fermi surface consists of isolated Dirac points, while those possessing type-II fermions are metals because of hole/electron pockets around the Fermi surface [11, 13-15]. Up to date, type-I Dirac fermions have been observed in lots of two- and three-dimensional materials [16-20]. However, type-II fermions are only found in a few heavy compounds, such as $PtSe_2$, $WTe_2$ and $MoTe_2$, hoping that the strong spin-orbit coupling (SOC) could help achieve a nontrivial band structure [11, 12]. It is naturally to ask if the type-II Dirac fermions can exist in the materials with negligible SOC, especially in graphene and its derivatives.

On the other hand, a number of functionalized methods have been adopted to tune graphene's electronic properties, such as doping and defects [21-25]. For the doping, N dopants are widely used, and N-doped graphene have been experimentally synthesized by chemical vapor deposition, thermal annealing graphite oxide with melamine and hyperthermal ion implantation, etc [26-28]. Various significant effects after N doping have been investigated [28-32]: band gap of graphene can be modulated by N delta-doping [32]; charge transfer, shift of the Fermi level and localized electronic state are found in N-doped graphene [31]. As to the defects, different defects in graphene have been studied theoretically and experimentally, such as Stone-Wales defects, 5-8 rings [33, 34]. Besides these localized defects, special line defects are also observed in graphene [21, 35-38]. A line defect, containing pentagonal and octagonal (5-8) rings embedded in a perfect graphene sheet, has been reported [34, 35, 39, 40]. The defect acts as a quasi-one-dimensional metallic wire, which may form building blocks for atomic-scale electronics [41, 42].

Here, we study electronic properties of graphene with nitrogen linear defects (NLD), using the first principles methods. By varying the widths $W$ of graphene nanoribbons between the NLD, a series of structures can be obtained. Interestingly, both type-I and type-II Dirac fermions are found in these line defect structures. When $W = 3n-1$ and $3n$ ($n$ is an integer), the structures have standard Dirac cones around the Fermi level, producing type-I Dirac fermions. When $W = 3n+1$, the structures have anisotropic tilted Dirac cones. The tilted Dirac cones cross the Fermi level and form hole/electron pockets, producing type-II Dirac fermions. The originations of the two types of Dirac fermions are explained by tight-binding models. All the Dirac fermions exist in the "bulk" rather than the line defects, which is different from the cases that defects induce Dirac fermions [43, 44]. Remarkably, there exist different edges states linking the two types of Dirac points. In the type-I and type-II Dirac materials, one and two Fermi arcs are formed between the type-I and type-II Dirac points, respectively.

## 2. Model and Computational Methods

Figure 1 shows the atomic structure of graphene embedding periodically NLDs, where the nitrogen molecules are implanted at the edges of armchair nanoribbons. Thus, the graphene nanoribbons are connected by 5-8 rings. The width of graphene nanoribbon between two line defects is labeled as $W$, and then the defect structure is named as NLD-graphene-$W$. When $W$ is changed, a series of structures can be obtained. Figures 1(a) and 1(b) present two kinds of NLD-graphene-$W$, where $W$ are even and odd, respectively. The dashed lines show the primitive cells of the two structures.

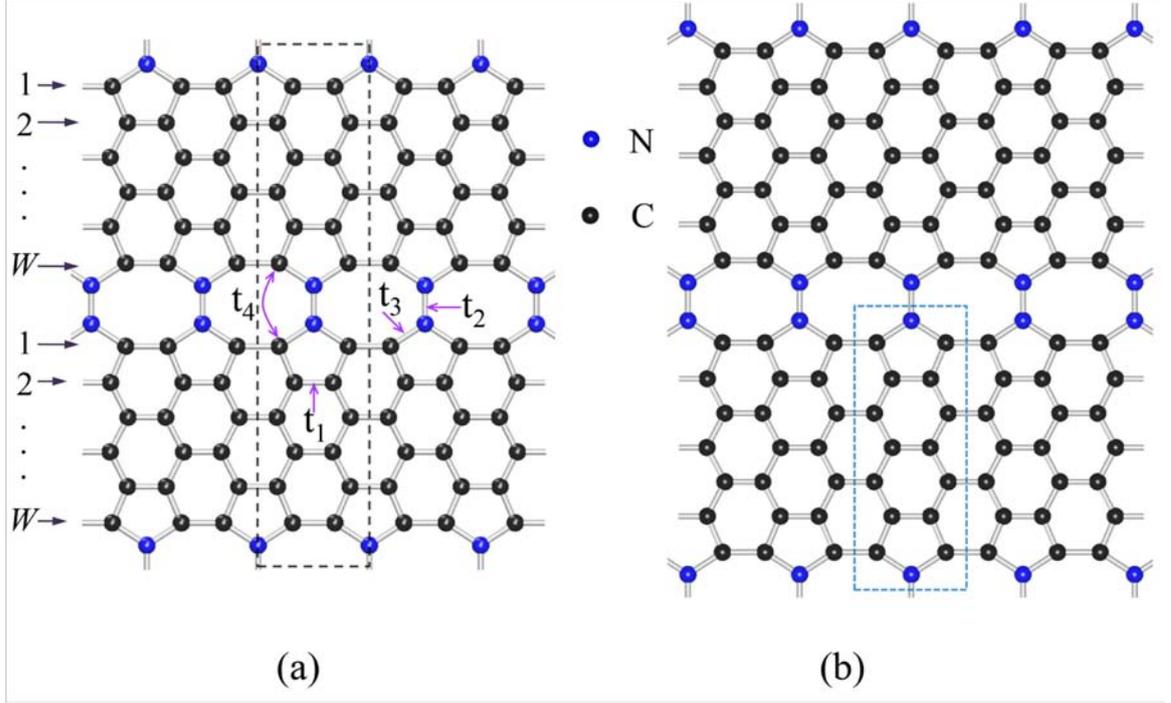

Figure 1. Atomic structures of NLD-graphene-$W$ with $W$ being an even number (a) and odd numbers (b), respectively, where $W$ represents the width of armchair graphene nanoribbons. Blue and black balls represent N and C atoms, respectively. The unit cells are shown in dashed lines. The $t_1$, $t_2$, $t_3$ and $t_4$ represent the hopping energies between different atoms.

We performed first-principles calculations based on the density functional theory (DFT) formalism as implemented in Vienna *ab initio* simulation program package (VASP) [45]. The Perdew-Burke-Ernzerhof (PBE) functional was employed for the exchange-correlation term according to generalized gradient approximation (GGA) [46, 47] . The projector augmented wave (PAW) method [48] has been used to represent the atomic cores. The kinetic energy cutoff of 500 eV was adopted. The atomic positions were optimized using the conjugate gradient method, the energy and force convergence criteria were set to be $10^{-5}$ eV and $10^{-2}$ eV/Å, respectively. A vacuum region of 10 Å was added to avoid interaction between adjacent images. A fine Monkhorst-Pack k-points meshes [49] were chosen to sampling the Brillion Zone (BZ)

## 3. Results and Discussions

In Fig. 2, the band structures of six NLD-graphene-$W$ structures are shown. It is known that the band structures especially band gaps of armchair graphene nanoribbons are dependent on their widths in a period of *three* [50, 51]. The band structures in Fig. 2 indicate that the electronic properties of NLD-graphene are also related to the width $W$. When $W = 3n-1$, a Dirac point appears on the Fermi level along M - Y and closes to the high symmetry point Y (see Figs. 2(a) and 2(b)). When $W = 3n$, there is a Dirac point along $\Gamma$- X and closing to $\Gamma$, as shown in Figs. 2(c) and 2(d). All the Dirac points in the two cases are crossed by two bands with inverse-sign slopes, which are called type-I Dirac points. The locations of the Dirac points in the first BZ for $W = 3n-1$ and $3n$ are shown in Figs. 3(a) and 3(b), respectively. According to the time reversal or inversion symmetries, each case has two Dirac points on the high-symmetry k paths parallel to the axis $k_x$. Figures 3(d) and 3(e) present the corresponding Dirac cones around the Dirac points, respectively. One can find that the Dirac cones are similar to the cones in the band structure of graphene. They are type-I Dirac cones and correspondingly produce type-I Dirac fermions. Therefore, type-I Dirac fermions exist in the NLD-graphene with $W = 3n-1$ and $3n$.

When $W = 3n+1$, there exist a crossing point along $\Gamma$- Y, as shown in Figs. 2(e) and 2(f). Different from the crossing points in the cases of $W = 3n-1$ and $3n$, the point here is crossed by two bands with same-sign slopes. One band has a larger slope while the other is close to a flat band. This kind of crossing point is called type-II Dirac point[11, 12] . According to the symmetry, there are two type-II Dirac points along the axis $k_y$ in the first BZ (see Figure 3(c)). In Fig. 3(f), we plot corresponding type-II Dirac cone around the points, which is obviously different from those in Figs. 3(d) and 3(e). It is a strong anisotropic cone with a tilted center axis. The cone crosses the Fermi level and thus forms hole and electron pockets. Then, two closed rings constitute the Fermi surface, as shown in Fig. 3(c), while the type-II Dirac points are contact points of the two rings. In this case, the line defect structures possessing type-II Dirac fermions are metals rather than semimetals.

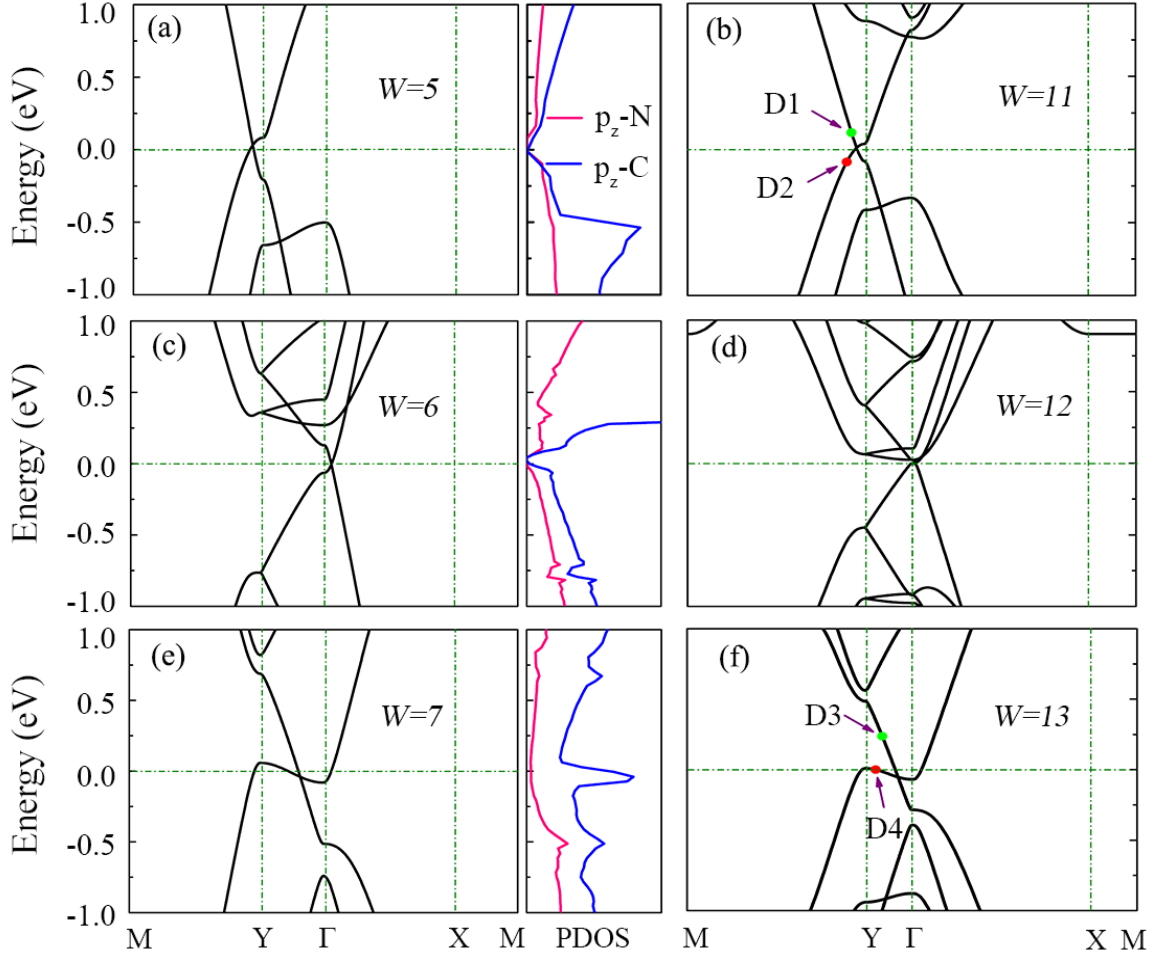

Figure 2. Band structures and PDOS of NLD-graphene-*W* with (a) *W=5*, (b) *W=11*, (c) *W=6*, (d) *W=12*, (e) *W=7* and (f) *W=13*, respectively.

As mentioned above, type-I and type-II Dirac points are crossings of different types of bands. To analyze the bands, in the right panels of Figs. 2(a), 2(c) and 2(e), we calculate the projected densities of states (PDOS) for the defect structures with $W = 3n-1$, $3n$ and $3n+1$, respectively. One can find that the band edge states around the Fermi level are all contributed by $p_z$ orbitals of C and N atoms. The type-I Dirac bands of $W = 3n-1$ and $3n$ are all attributed to the common contributions of N and C atoms. However, the case of type-II Dirac bands is different: the large-slope band is attributed to both C and N atoms, while the small-slope band is only attributed to C atoms. To display the different characteristics of the two types of bands clearly, in Fig. 4, we compare their charge densities of states. Figure 4(a) presents two states (D1 and D2) around the type-I Dirac point in Fig. 2(b), while Fig. 4(b) presents two states (D3 and D4) around the type-II Dirac point in Fig. 2(f). In the states D1, D2 and D3, the electron distribute in the "bulk" including N and C atoms. However, in the state D4, no electrons distribute on the N atoms, instead the electrons are only "localized" in the graphene nanoribbons. In another words, the state D4 is a quasi-one-dimensional state rather than a "bulk" state. We note that the states D1, D2 and D3 corresponds to the large-slope Dirac bands, while the state D4 corresponds to the small-slope Dirac band. This implies that formation of the type-II Dirac point is related to the quasi-one-dimensional (Q1D) "localized states" in the graphene nanoribbons.

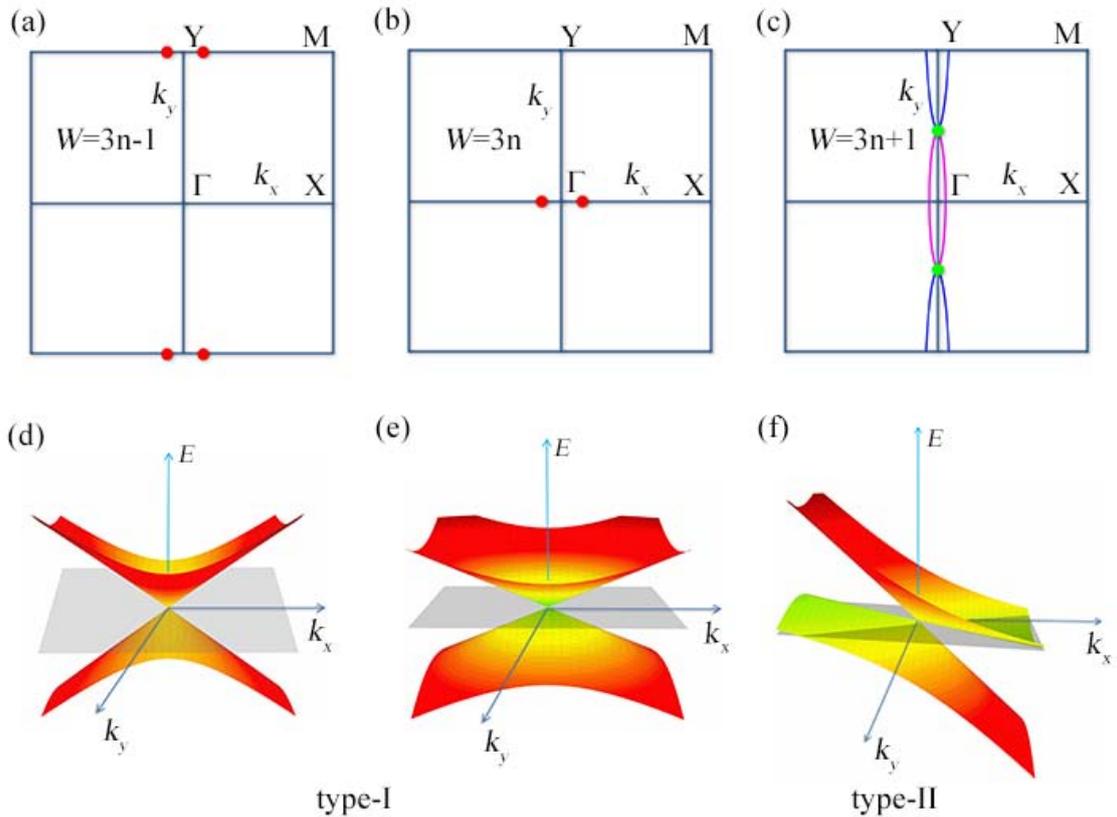

Figure 3. The first BZ of NLD-graphene-*W* with (a) *W=3n-1*, (b) *W=3n*, (c) *W=3n+1*. The red dots in (a) and (b) represent locations of the type-I Dirac points, while the green dots in (c) represent locations of the type-II Dirac points. The blue and pink solid lines in (c) represent crossing rings of hole/electron pockets and the Fermi level. (d-e) Type-I Dirac cones in NLD-graphene-*W* with *W=3n-1* and *3n*, respectively. (f) Type-II Dirac cone in NLD-graphene-*W* with *W=3n+1*. The gray planes in (d-f) represent the Fermi level. One can clearly find that the type-II Dirac cone cross the Fermi level and then forms hole and electron pockets.

To further explore originations of the two types of Dirac bands, we use a tight-binding model to explain evolution of band structures with orbital interactions. Because only p$_z$ orbital has contributions to the bands around the Fermi level, a single orbital tight-binding Hamiltonian is adopted to describe electronic properties of the defect structures,

$$H = (\varepsilon_C \sum_i a_i^\dagger a_i + t_1 \sum_{ij} a_i^\dagger a_j) + (\varepsilon_N \sum_l b_l^\dagger b_l + t_2 \sum_{lm} b_l^\dagger b_m)$$
$$+ t_3 \sum_{il} a_i^\dagger b_l + t_4 \sum_{ij'} a_i^\dagger a_{j'}, \quad (1)$$

where $a_i^\dagger/b_l^\dagger$ and $a_j/b_m$ represent the creation/annihilation operators, the first two terms are Hamiltonians of graphene nanoribbons and N molecules, respectively, the third term represents interaction between the nanoribbons and N molecules, while the last term represents the interaction between the graphene nanoribbons. $\varepsilon_C$ and $\varepsilon_N$ are site energies of C and N atoms, $t_1$, $t_2$ and $t_3$ are hopping energies of C atoms in graphene nanoribbons, N atoms and between them, respectively, and $t_4$ is hopping energy between C atoms in two neighboring graphene nanoribbons. All hopping energies are illustrated in Fig. 1(a).

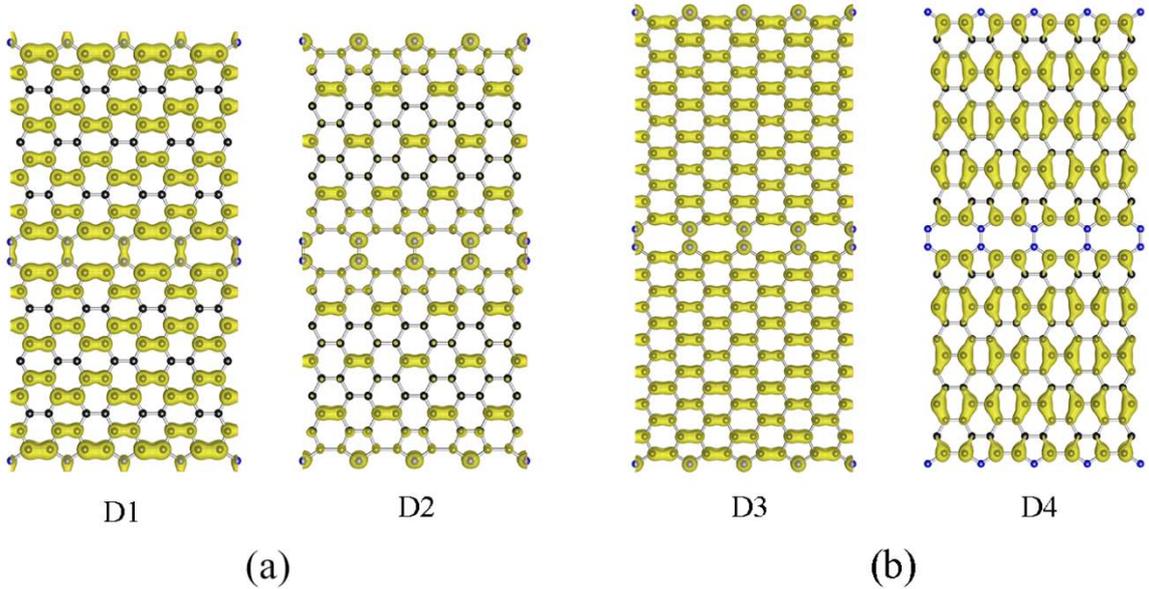

Figure 4. (a) Charge densities of states at D1 and D2 in Fig. 2(b), respectively. (b) Charge densities of states at D3 and D4 in Fig. 2(f), respectively.

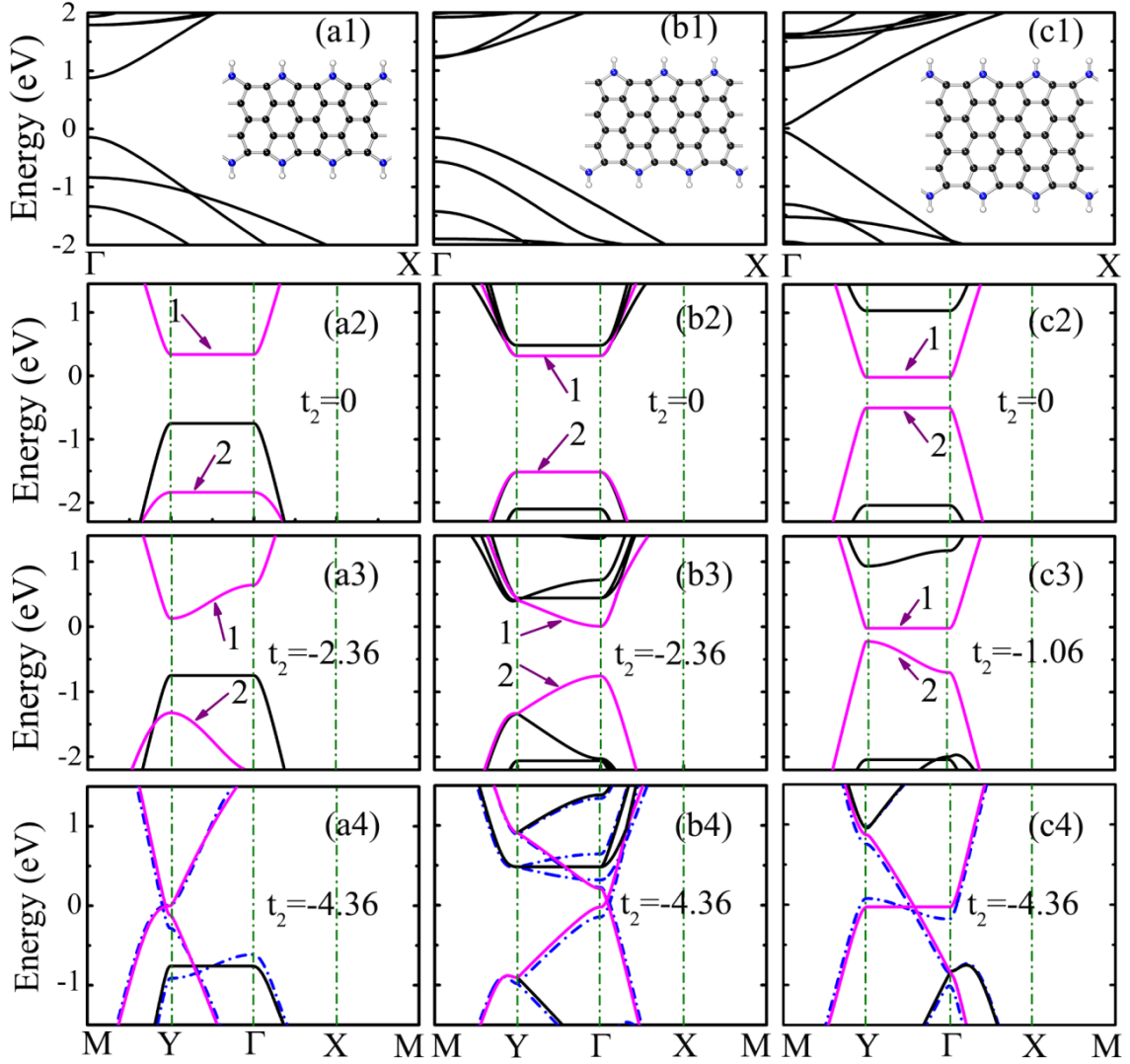

Figure 5. (a1)-(c1) DFT band structures of N-edged armchair graphene nanoribbons with widths $W = 5, 6, 7$. The atomic structures of the N-edged armchair graphene nanoribbons are shown in the insets. Tight-binding band structures of NDL-graphene-$W$ with (a2)-(a4) $W=5$, (b2)-(b4) $W=6$ and (c2)-(c4) $W=7$ by varying the hopping energy $t_2$, respectively. The other parameter are $t_1=-2.36$ eV, $t_3= -3.12$ eV and $t_4= 0$ eV. When $t_4=-0.3$ eV, the tight-binding band structures are marked as blue dashed lines in (a4), (b4) and (c4).

The evolutions of tight-binding band structures with $t_2$ for NLD-graphene-*5*, *6* and *7* are shown in Figs. 5(a2-c4). When $t_2 = 0$, it means that there are no interactions between the N-edged nanoribbons, and thus the band structures along $\Gamma$ - X for the line defect structures should be similar to those of N-edged nanoribbons. For comparison, in Figs. 5(a1-c1), we present band structures of the N-edged nanoribbons with $W = 5, 6$, and *7* by DFT calculations, respectively. Comparing Figs.

5(a1-c1) with Figs. 5(a2-c2), one can find that the tight-binding results fit well with the DFT results. The band gaps of *W = 5* and *6* are larger than that of *W =7*. When $t_2 > 0$, the interaction between N-edged nanoribbons increases with $t_2$. Seen from Figs. 5(a3-b4), for the NLD-graphene-*5* and *6*, the bands labeled as 1 and 2 gradually close together with the increase of $t_2$. After crossing, band inversion induces the formation of type-I Dirac points. While the band evolution of NLD-graphene-*7* is different, as shown in Figs. 5(c3-c4). The band 2 is very sensitive to the variation of $t_2$, while band 1 is insensitive and always flat along Γ - Y. The insensitivity of band 1 further proves that band 1 is induced by the Q1D "localized states" in graphene nanoribbons. After the two bands cross, band inversion induces a tilted type-II Dirac cone.

The former discussion about the band evolution with $t_2$ is based on the case of $t_4 = 0$, i.e., we do not consider the interactions between graphene nanoribbons. As $t_4 \neq 0$, it nearly has no effect to the type-I Dirac bands (see the dashed lines in Figs. 5(a4) and 5(b4)), because these bands are mainly affected by $t_2$. However, $t_4$ has a significant effect on the flat band in Fig. 5(c4). The flat band becomes a linear band with a small slope, which is more close to the DFT results in Fig. 2(e). Therefore, one can conclude that the type-II Dirac bands are originated from weak interactions between the Q1D "localized states" in the graphene nanoribbons.

In Fig. 6, we calculate edge states of the two types of Dirac materials. Figures 6(a) and 6(b) present band structures of NLD-graphene-*5*, *6* after cutting along armchair direction, respectively. The two 2D structures possess type-I Dirac points on the k path parallel to the axis $k_x$. These Dirac points are projected into the band structures of 1D nanoribbons, which is labeled as red circles in Figs. 6(a) and 6(b). There exists a obvious Fermi arc between the Dirac points. This is somewhat similar to the case of graphene. However, the edge states in these defect structures are along armchair direction, while those in graphene are along zigzag direction. Figure 6(c) presents band structure of NLD-graphene-*7* after cutting along zigzag direction. The type-II Dirac points are also projected into the band structure of 1D nanoribbon (see the green circles in Fig. 6(c)). It is noted that there exist two Fermi arcs between the Dirac points. One of them is merged into the bulk states while the other is clearly seen below the Fermi level.

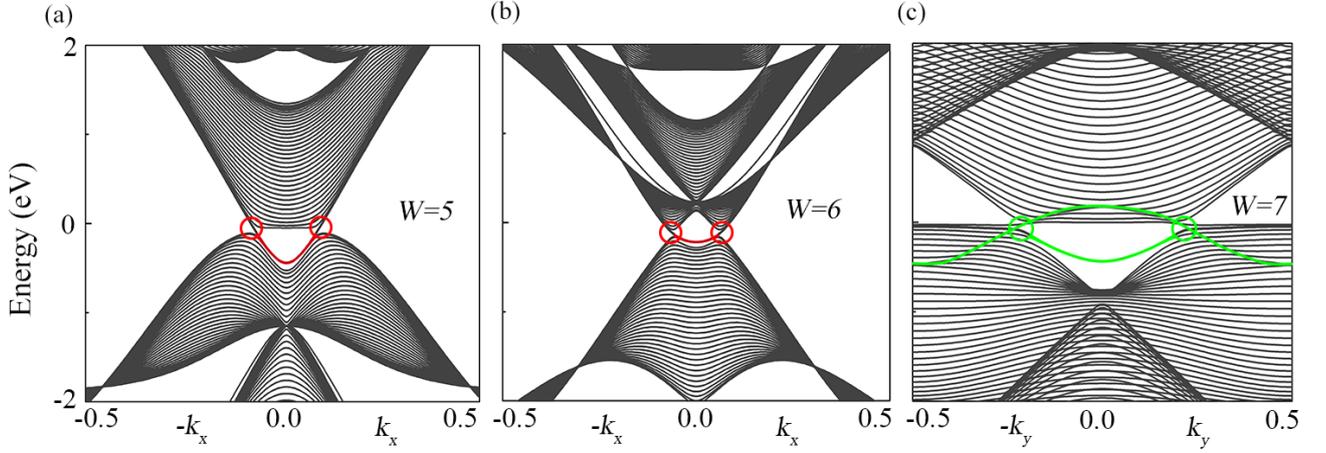

Figure 6. Band structures for nanoribbons of NLD-graphene-$W$ with (a) $W=5$ and (b) $W=6$ along armchair direction, with (c) $W=7$ along zigzag direction. The red circles in (a) and (b) are projections of the type-I Dirac points, and the red lines are Fermi arcs. The green circles in (c) are projections of the type-II Dirac points, and the green lines are Fermi arcs between the points.

## 4. Conclusions

We have studied electronic properties of NLD-graphene, by using the first-principles method and tight-binding model. The line defect structures exhibit diverse band structures, depending on the size $W$ of the armchair-edged nanoribbons between line defects. There are standard type-I Dirac bands around the Fermi level when $W = 3n-1$ and $3n$. When $W = 3n+1$, type-II Dirac fermions are found, with strongly tilted Dirac cones. The analysis of PDOS and a tight-binding model explains the origination of the different Dirac fermions. The two types of Dirac points leads to two kinds of edge states: the type-I Dirac points induce one Fermi arc corresponding to edge states along the armchair direction, while the type-II Dirac points induce two Fermi arcs corresponding to two sets of edge states along the zigzag direction.

It is noted that this is the first time that type-II Dirac fermions are found in the light-element structures with negligible SOC. Moreover, these type-II Dirac points are easy to be observed in experiment, because no any other bands cross the Fermi level. Our studies provide a good supporter not only to find type-II Dirac fermions but also to compare the difference between type-I and type-II Dirac fermions. This expands our views on the Dirac fermions in two-dimensional structures, and also extend their applications in electronics.

## Acknowledgments

This work was supported by the National Natural Science Foundation of China (Nos. 51376005 and 11474243).

# References

[1] Castro Neto AH, Guinea F, Peres NMR, Novoselov KS, Geim AK. The electronic properties of graphene. Rev Mod Phys. 2009;81(1):109-62.

[2] Novoselov KS. Nobel Lecture: Graphene: Materials in the Flatland*. Rev Mod Phys. 2011;83(3):837-49.

[3] Young AF, Sanchez-Yamagishi JD, Hunt B, Choi SH, Watanabe K, Taniguchi T, et al. Tunable symmetry breaking and helical edge transport in a graphene quantum spin Hall state. Nature. 2014;505(7484):528-32.

[4] Du X, Skachko I, Duerr F, Luican A, Andrei EY. Fractional quantum Hall effect and insulating phase of Dirac electrons in graphene. Nature. 2009;462(7270):192-5.

[5] Geim AK. Graphene: status and prospects. Science. 2009;324(5934):1530-4.

[6] Novoselov KS, Geim AK, Morozov SV, Jiang D, Katsnelson MI, Grigorieva IV, et al. Two-dimensional gas of massless Dirac fermions in graphene. Nature. 2005;438(7065):197-200.

[7] Zhang Y, Tan YW, Stormer HL, Kim P. Experimental observation of the quantum Hall effect and Berry's phase in graphene. Nature. 2005;438(7065):201-4.

[8] Bolotin KI, Sikes K, Jiang Z, Klima M, Fudenberg G, Hone J, et al. Ultrahigh electron mobility in suspended graphene. Solid State Commun. 2008;146(9):351-5.

[9] Hwang E, Sarma SD. Acoustic phonon scattering limited carrier mobility in two-dimensional extrinsic graphene. Phys Rev B. 2008;77(11):115449.

[10] Andrei Bernevig B, Hughes TL, Zhang S-C, Chen H-D, Wu C. BAND COLLAPSE AND THE QUANTUM HALL EFFECT IN GRAPHENE. Int J Mod Phys B. 2006;20(22):3257-78.

[11] Huang H, Zhou S, Duan W. Type-II Dirac fermions in the PtSe2 class of transition metal dichalcogenides. Phys Rev B. 2016;94(12):121117.

[12] L Muechler AA, T Neupert. Topological metals from band inversion. arXiv preprint arXiv:160401398. 2016.

[13] M Yan HH, K Zhang, E Wang, W Yao. Lorentz-violating type-II Dirac fermions in transition metal

dichalcogenide PtTe2. arXiv preprint arXiv:160703643. 2016.

[14] TR Chang SX, DS Sanchez, SM Huang. Type-II Topological Dirac Semimetals: Theory and Materials

Prediction (VAl3 family). arXiv preprint arXiv:160607555. 2016.

[15] CK Chiu YC, X Li, Y Nohara. Type-II Dirac surface states in topological crystalline insulators. arXiv preprint arXiv:160603456. 2016.

[16] Wehling TO, Black-Schaffer AM, Balatsky AV. Dirac materials. Adv Phys. 2014;63(1):1-76.

[17] Zhong C, Xie Y, Chen Y, Zhang S. Coexistence of flat bands and Dirac bands in a carbon-Kagome-lattice family. Carbon. 2016;99:65-70.

[18] Chen Y, Xie Y, Yang SA, Pan H, Zhang F, Cohen ML, et al. Nanostructured Carbon Allotropes with Weyl-like Loops and Points. Nano Lett. 2015;15(10):6974-8.

[19] Wang Z, Zhou XF, Zhang X, Zhu Q, Dong H, Zhao M, et al. Phagraphene: A Low-Energy Graphene Allotrope Composed of 5-6-7 Carbon Rings with Distorted

Dirac Cones. Nano Lett. 2015;15(9):6182-6.
[20] Wang J, Deng S, Liu Z, Liu Z. The rare two-dimensional materials with Dirac cones. Natl Sci Rev. 2015;2(1):22-39.
[21] Vicarelli L, Heerema SJ, Dekker C, Zandbergen HW. Controlling defects in graphene for optimizing the electrical properties of graphene nanodevices. ACS nano. 2015;9(4):3428-35.
[22] Lherbier A, Liang L, Charlier J-C, Meunier V. Charge carrier transport and separation in pristine and nitrogen-doped graphene nanowiggle heterostructures. Carbon. 2015;95:833-42.
[23] Li Y, Zhou Z, Shen P, Chen Z. Spin Gapless Semiconductor−Metal−Half-Metal Properties in Nitrogen-Doped Zigzag Graphene Nanoribbons. ACS Nano. 2009;3(7):1952-8.
[24] Kang J, Bang J, Ryu B, Chang KJ. Effect of atomic-scale defects on the low-energy electronic structure of graphene: Perturbation theory and local-density-functional calculations. Phys Rev B. 2008;77(11).
[25] Wang Z, Zhou YG, Bang J, Prange MP, Zhang SB, Gao F. Modification of Defect Structures in Graphene by Electron Irradiation: Ab Initio Molecular Dynamics Simulations. J Physl Chem C. 2012;116(30):16070-9.
[26] Cress CD, Schmucker SW, Friedman AL, Dev P, Culbertson JC, Lyding JW, et al. Nitrogen-Doped Graphene and Twisted Bilayer Graphene via Hyperthermal Ion Implantation with Depth Control. ACS Nano. 2016;10(3):3714-22.
[27] Sheng Z-H, Shao L, Chen J-J, Bao W-J, Wang F-B, Xia X-H. Catalyst-Free Synthesis of Nitrogen-Doped Graphene via Thermal Annealing Graphite Oxide with Melamine and Its Excellent Electrocatalysis. ACS Nano. 2011;5(6):4350-8.
[28] Wei D, Liu Y, Wang Y, Zhang H, Huang L, Yu G. Synthesis of N-Doped Graphene by Chemical Vapor Deposition and Its Electrical Properties. Nano lett. 2009;9(5):1752-8.
[29] Schiros T, Nordlund D, Pálová L, Prezzi D, Zhao L, Kim KS, et al. Connecting dopant bond type with electronic structure in N-doped graphene. Nano lett. 2012;12(8):4025-31.
[30] Sforzini J, Hapala P, Franke M, van Straaten G, Stöhr A, Link S, et al. Structural and Electronic Properties of Nitrogen-Doped Graphene. Phys Rev Lett. 2016;116(12):126805.
[31] Zheng B, Hermet P, Henrard L. Scanning tunneling microscopy simulations of nitrogen-and boron-doped graphene and single-walled carbon nanotubes. Acs Nano. 2010;4(7):4165-73.
[32] Wei X-L, Fang H, Wang R-Z, Chen Y-P, Zhong J-X. Energy gaps in nitrogen delta-doping graphene: a first-principles study. Appl Phys Lett. 2011;99(1):012107.
[33] Ma J, Alfe D, Michaelides A, Wang E. Stone-Wales defects in graphene and other planar s p 2-bonded materials. Phys Rev B. 2009;80(3):033407.
[34] Lahiri J, Lin Y, Bozkurt P, Oleynik II, Batzill M. An extended defect in graphene as a metallic wire. Nat nanotechnol. 2010;5(5):326-9.
[35] Alexandre SS, Lúcio A, Neto AC, Nunes R. Correlated magnetic states in extended one-dimensional defects in graphene. Nano lett. 2012;12(10):5097-102.

[36] Kou L, Tang C, Guo W, Chen C. Tunable magnetism in strained graphene with topological line defect. ACS nano. 2011;5(2):1012-7.
[37] Okada S, Kawai T, Nakada K. Electronic Structure of Graphene with a Topological Line Defect. J Phys Soc Jan. 2011;80(1):013709.
[38] Gunlycke D, White CT. Graphene valley filter using a line defect. Phys Rev Lett. 2011;106(13):136806.
[39] Botello-Méndez AR, Declerck X, Terrones M, Terrones H, Charlier J-C. One-dimensional extended lines of divacancy defects in graphene. Nanoscale. 2011;3(7):2868-72.
[40] Song J, Liu H, Jiang H, Sun Q-f, Xie X. One-dimensional quantum channel in a graphene line defect. Phys Rev B. 2012;86(8):085437.
[41] Li Y, Zhang R-Q, Lin Z, Van Hove MA. Inducing extended line defects in graphene by linear adsorption of C and N atoms. Appl Phys Lett. 2012;101(25):253105.
[42] Ren J-C, Ding Z, Zhang R-Q, Van Hove MA. Self-doping and magnetic ordering induced by extended line defects in graphene. Phys Rev B. 2015;91(4):045425.
[43] Yu S, Zheng W, Ao Z, Li S. Confinement of massless Dirac fermions in the graphene matrix induced by the B/N heteroatoms. Phys Chem Chem Phys. 2015;17(8):5586-93.
[44] Lima EN, Schmidt TM, Nunes RW. Topologically Protected Metallic States Induced by a One-Dimensional Extended Defect in the Bulk of a 2D Topological Insulator. Nano Lett. 2016;16(7):4025-31.
[45] Kresse G, Hafner J. Ab initio molecular dynamics for liquid metals. Phys Rev B. 1993;47(1):558.
[46] Perdew JP, Burke K, Ernzerhof M. Generalized Gradient Approximation Made Simple. Phys Rev Lett. 1996;77(18):3865-8.
[47] Perdew JP, Chevary J, Vosko S, Jackson KA, Pederson MR, Singh D, et al. Atoms, molecules, solids, and surfaces: Applications of the generalized gradient approximation for exchange and correlation. Phys Rev B. 1992;46(11):6671.
[48] Kresse G, Joubert D. From ultrasoft pseudopotentials to the projector augmented-wave method. Phys Rev B. 1999;59(3):1758.
[49] Monkhorst HJ, Pack JD. Special points for Brillouin-zone integrations. Phys Rev B. 1976;13(12):5188-92.
[50] Son YW, Cohen ML, Louie SG. Energy gaps in graphene nanoribbons. Phys Rev Lett. 2006;97(21):216803.
[51] Ma F, Guo Z, Xu K, Chu PK. First-principle study of energy band structure of armchair graphene nanoribbons. Solid State Commun. 2012;152(13):1089-93.